\documentclass[pra,aps,floatfix,showpacs,tightenlines,twocolumn,
amsmath,amssymb]{revtex4}

\usepackage{graphicx}
\usepackage{dcolumn}
\usepackage{bm}
\usepackage{epsfig}

\begin{document}

\title{Quantum Repeaters using Coherent-State Communication}

\author{Peter van Loock$^{1,2}$, Norbert L\"utkenhaus$^{3}$,
W. J. Munro$^{2,4}$, and Kae Nemoto$^{2}$}

\affiliation{$^{1}$Optical Quantum Information Theory Group,
Institute of Theoretical Physics I and Max-Planck Research Group,
Institute of Optics, Information and Photonics,
Universit\"{a}t Erlangen-N\"{u}rnberg, Staudtstr. 7/B2,
91058 Erlangen, Germany\\
$^{2}$National Institute of
Informatics, 2-1-2 Hitotsubashi, Chiyoda-ku, Tokyo 101-8430,
Japan\\
$^{3}$Institute for Quantum Computing, University of Waterloo, Canada\\
$^{4}$Hewlett-Packard Laboratories, Filton Road, Stoke
Gifford, Bristol BS34 8QZ, United Kingdom}

\begin{abstract}
We investigate quantum repeater protocols based upon atomic qubit-entanglement
distribution through optical coherent-state communication. Various measurement
schemes for an optical mode entangled with two spatially separated atomic
qubits are considered in order to nonlocally prepare
conditional two-qubit entangled states. In particular, generalized measurements
for unambiguous state discrimination enable one to completely eliminate spin-flip
errors in the resulting qubit states, as they would occur in a homodyne-based
scheme due to the finite overlap of the optical states in phase space.
As a result, by using weaker coherent states, high initial fidelities can still be achieved
for larger repeater spacing, at the expense of lower entanglement generation rates. In this regime,
the coherent-state-based protocols start resembling single-photon-based repeater schemes.
\end{abstract}


\maketitle

\section{Introduction}

In long-distance, classical communication networks, signals that are gradually
distorted during their propagation through a channel are repeatedly recreated via
a chain of intermediate stations along the transmission line.
For instance, optical pulses traveling through a glass fiber and being subject to
photon loss can be reamplified at each repeater station. Such an amplification is
impossible, when the signal carries quantum information. If a quantum bit is encoded
into a single photon, its unknown quantum state cannot be copied along the line
\cite{WoottersZurek,Dieks}; the photon must travel the entire distance with an exponentially
decreasing probability to reach the end of the channel.

The solution to the problem of long-distance quantum communication
is provided by the so-called quantum repeater \cite{Briegel98,Duer99}.
In this case, prior to the actual quantum-state communication, a supply of
standard entangled states is generated and distributed among not too distant
nodes of the channel. If sufficiently many of these imperfect entangled states
are shared between the repeater stations, a combination of entanglement
purification and entanglement swapping extends this shared entanglement over
the entire channel. Through entanglement swapping \cite{Zukowski},
the entanglement of neighboring pairs is connected,
gradually increasing the distance of the shared
entanglement. The entanglement purification \cite{Bennett96,Deutsch}
enables one to distill (through local operations) a high-fidelity entangled pair
from a larger number of low-fidelity entangled pairs,
as they would emerge after a few rounds of entanglement swapping with imperfect
entangled states, or even at the very beginning after the initial, imperfect
entanglement generation and distribution.

Current implementations for quantum communication, in particular,
quantum key distribution, are limited by a distance
of about 200 km. In principle, one could go beyond this distance using
a quantum repeater. However, the issue of actually realizing a quantum repeater
protocol is rather subtle, even for not too long distances. In particular,
the subroutines of entanglement distillation and swapping require advanced local
quantum logic including, for instance, two-qubit entangling gates; moreover, a sufficient
quantum memory is needed such that local measurement results can be communicated between
the repeater stations \cite{Hartmann07}.
Nonetheless, various proposals exist, of which the most recent ones
are based on the nonlocal generation of atomic (spin) entangled states, conditioned upon
the detection of photons distributed between two neighboring repeater stations.
The light, before traveling through the communication channel and being detected,
is scattered from either individual atoms, for example, in form of solid-state single photon
emitters \cite{Childress1,Childress2}, or from an atomic ensemble,
i.e. a cloud of atoms in a gas \cite{Duan01}. In these heralded schemes, typically,
the fidelities of the initial entanglement generation are quite high, at the expense of
rather small efficiencies. Other complications include interferometric phase stabilization
over large distances \cite{Simon07,Zhao07,Chen07} and the purification of atomic ensembles. Yet some elements towards
a realization of the protocol in Ref.~\cite{Duan01} have been demonstrated already
\cite{Chou05,Chaneliere05,Eisaman05}. Further theoretical results were presented very recently
\cite{Jiang07}, improving the scheme of Ref.~\cite{Duan01}.

In this paper, we will extend our previous results on the so-called hybrid quantum repeater
\cite{vanloock06,Ladd06}. This approach to long-distance quantum communication is somewhat different
from those mentioned above. It relies on atom-light entanglement which becomes manifest
in quantum correlations between a discrete spin variable and a continuous optical phase quadrature
rather than a discrete single-photon occupation number. An optical pulse in a coherent state of about
$10^{4}$ photons is subject to a controlled phase rotation (achieved through dispersive, CQED-type
interactions), conditioned upon the state of the atom. After propagating to the nearest neighboring repeater station
and a further interaction with a second spin at that station,
the light field is measured via homodyne detection and an imperfect entangled two-qubit state is nonlocally
prepared between the two repeater stations through postselection. Finally, the same
dispersive light-matter interactions are exploited to achieve the local quantum gates
(``qubus computation'' \cite{Spiller06,vanloock07}) needed for entanglement purification and swapping.

The two main advantages of the hybrid repeater protocol, distinct from the single-photon-based
schemes, are the high success probabilities for postselection in the entanglement generation step
and the intrinsic phase stabilization provided through reference pulses propagating in the same channel
as the probes. However, these assets are at the expense of rather modest initial fidelities of the entangles states
and high sensitivity to photon losses and noise in the optical channel. In fact, distances between repeater stations
beyond 10-20 km turn out to be impossible with the current proposal; the
decoherence effect (a damping of the off-diagonal terms of the two-qubit density matrix after postselection)
exponentially grows with distance such that only smaller mean photon numbers lead to a sufficient degree
of entanglement; however, the less intense coherent states are less distinguishable, hence resulting in
a further decrease of fidelity through postselection errors. A good trade-off between these competing
sources of errors is only possible for not too large distances.

The analysis here will provide a possible solution to the distance limitation. This is particularly
important, as the typical repeater spacing in existing classical communication networks is of the order of 50-100 km
and thus incompatible with the current hybrid repeater protocol.
The distance limit can be overcome by completely eliminating one source of errors, namely that which stems from
the finite overlaps of the phase-rotated coherent states. This is achieved through a different detection scheme,
where the coherent states are unambiguously discriminated. Such an unambiguous state discrimination (USD) is error-free;
so for nonorthogonal states, it must include inconclusive measurement results. These will lead to lower efficiencies
of the entanglement generation, in particular, when smaller photon numbers are used in order to
suppress the decoherence effect through photon losses. The corresponding trade-off between success probability and fidelity
means there are ultimate quantum mechanical bounds on the accessible regimes. We will discuss these bounds and
propose suboptimal, but practical, linear optical implementations.

The emphasis here is on possible measurement schemes
for the initial entanglement generation. Further, we investigate the hybrid entangled atom-light states before the
measurements and potential variations of the entanglement distillation and swapping steps. The latter could be performed
already on the atom-light level
(``hybrid entanglement distillation and swapping'')
rather than solely on the atomic level after the conditional state preparation. We do not consider issues related with the CQED
part (for this, see \cite{Ladd06}); neither are we concerned about architecture-related issues on how to
combine the entanglement purification and swapping steps in an optimal way (for this, see \cite{vanmeter07}).
Such considerations will be needed for comparing the overall efficiencies between the hybrid approach
and the single-photon-based schemes.

The plan of the paper is as follows. First, in
Sec.~\ref{entgen}, we will examine the hybrid entangled states
between one atomic spin and an optical mode (Sec.~\ref{entgen1}),
the entangled states of two spins and an optical mode (Sec.~\ref{entgen2}),
and the measurements for conditional entangled-state preparation
(Sec.~\ref{entgen3}). Secondly, in Sec.~\ref{entpurif},
we will discuss the notions of hybrid entanglement distillation and swapping
and their potential realizations.

\section{Entanglement Generation}\label{entgen}

\begin{figure}[b]
\includegraphics[width=\columnwidth]{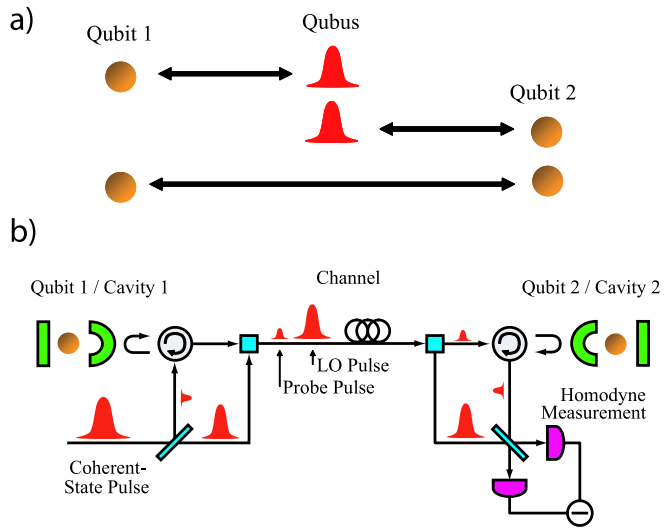}
\caption{\label{fig1} a) Three steps for the generation of
spin-entanglement between two qubits at neighboring repeater stations:
the first interaction results in an entangled state between the atomic
qubit 1 and the optical qubus; after transmission, the qubus interacts with
the atomic qubit 2, leading to a tripartite entangled state between
the two qubits and the qubus; finally, a measurement on the qubus mode
conditionally creates the two-qubit entanglement.
b) Example of a possible measurement scheme for discriminating between the conditionally
phase-rotated coherent probe beams in the hybrid quantum repeater;
the LO pulse is a sufficiently
strong local oscillator used for homodyne detection \cite{vanloock06}.}
\end{figure}

In the hybrid quantum repeater,
the mechanism for entanglement distribution is based
on dispersive light-matter interactions, obtainable
from the Jaynes-Cummings interaction Hamiltonian
$\hbar g (\hat\sigma^{-}\hat a^{\dagger} + \hat\sigma^{+}\hat a)$
in the limit
of large detuning \cite{Schleich},
\begin{equation}\label{Hint}
\hat H_{\rm int} = \hbar \chi \hat\sigma_z \hat a^{\dagger} \hat a\,.
\end{equation}
Here, $\hat a$ ($\hat a^{\dagger}$) is the annihilation (creation)
operator of the electromagnetic field mode and
$\hat\sigma_z = |0\rangle\langle 0| - |1\rangle\langle 1|$ is the
corresponding qubit Pauli operator for a two-level atom.
The parameter $\chi=g^2/\Delta$ describes the strength of the atom-light
coupling; $2g$ is
the vacuum Rabi splitting for the dipole transition and $\Delta$
is the detuning between the dipole transition and the light field.
The Hamiltonian in Eq.~(\ref{Hint}) leads to a conditional
phase-rotation of the field mode,
\begin{eqnarray}\label{interaction1}
\hat U_{\rm int} = \exp(i\theta \hat \sigma_z \hat a^\dagger \hat a
)\,.
\end{eqnarray}
Here, $\chi t\equiv \theta$ is an effective interaction time.
The only requirement for a dispersive interaction resulting in a
high-fidelity conditional rotation is a sufficiently large
cooperativity parameter in a weak or intermediate coupling regime; strong
coupling is not needed \cite{Ladd06}.
For simplicity,
let us now write the effect of a controlled rotation on a coherent state
and a qubit superposition state (corresponding to Eq.~(\ref{interaction1})
up to an uncontrolled phase rotation) as
\begin{eqnarray}\label{interaction2}
\hat U_{\rm int} \left[\left( |0\rangle+|1\rangle \right)
|\alpha\rangle\right]/\sqrt{2} = \left(|0\rangle
|\alpha\rangle+|1\rangle |\alpha e^{i
\theta}\rangle\right)/\sqrt{2}\,.
\end{eqnarray}
In the following, we will investigate the entangled states that
emerge from this interaction. According to the initial entanglement distribution
procedure for the hybrid quantum repeater, as a first step, one atomic qubit
interacts with the optical qubus mode (see Fig.~\ref{fig1}a), resulting in a
``hybrid entangled state'' between qubit 1 and the qubus, as described
by Eq.~(\ref{interaction2}). During the transmission
of the qubus through the (lossy) channel, the hybrid entangled state is subject to decoherence
and becomes mixed.
Then the qubus interacts with qubit 2; at this stage, the two qubits
and the qubus are in a tripartite (mixed) entangled state.
Finally, a measurement on the qubus mode conditionally prepares a two-qubit
(mixed) entangled state. An example of a possible measurement scheme to (approximately)
achieve this final step is through homodyne detection (see Fig.~\ref{fig1}b).
Let us now closely examine the elements of this protocol with respect to possible
improvements.

\subsection{Qubit-qubus entanglement}\label{entgen1}

Let us assume the qubus mode is entangled with qubit 1
after the first interaction, as expressed by Eq.~(\ref{interaction2}).
Photon losses in the qubus channel are now described via a simple
beam splitter which reflects, on average, $1-\eta$ photons into an environment
mode, initially in the vacuum state $|0\rangle_{\rm E}$,
\begin{eqnarray}\label{losses}
&&|0\rangle_{\rm A}
|\sqrt{\eta}\alpha\rangle_{\rm B}
|\sqrt{1-\eta}\alpha\rangle_{\rm E}/\sqrt{2}
\nonumber\\
&&\quad+
|1\rangle_{\rm A}
|\sqrt{\eta}\alpha e^{i
\theta}\rangle_{\rm B}
|\sqrt{1-\eta}\alpha e^{i
\theta}\rangle_{\rm E}/\sqrt{2}\,.
\end{eqnarray}
The subscripts ``A'' and ``B'' denote the state of the atomic qubit and the
qubus mode, respectively. We may now rewrite each of the two pairs of pure, nonorthogonal
states of the qubus mode and the loss mode in an orthogonal, two-dimensional basis,
$\{|u\rangle,|v\rangle\}$,
\begin{eqnarray}\label{qubusbasis}
|\sqrt{\eta}\alpha\rangle_{\rm B}&=&\mu_{\rm B}|u\rangle_{\rm B}+
\nu_{\rm B}|v\rangle_{\rm B},\nonumber\\
|\sqrt{\eta}\alpha e^{i
\theta}\rangle_{\rm B}&=&(\mu_{\rm B}|u\rangle_{\rm B}-
\nu_{\rm B}|v\rangle_{\rm B})
\,e^{i\eta\xi},\nonumber\\
|\sqrt{1-\eta}\alpha\rangle_{\rm E}&=&\mu_{\rm E}|u\rangle_{\rm E}+
\nu_{\rm E}|v\rangle_{\rm E},\\
|\sqrt{1-\eta}\alpha e^{i
\theta}\rangle_{\rm E}&=&(\mu_{\rm E}|u\rangle_{\rm E}-
\nu_{\rm E}|v\rangle_{\rm E})
\,e^{i(1-\eta)\xi}\,,\nonumber
\end{eqnarray}
where $\nu_{\rm B}=\sqrt{1-\mu^2_{\rm B}}$ and
$\nu_{\rm E}=\sqrt{1-\mu^2_{\rm E}}$ with
\begin{eqnarray}\label{qubusbasis2a}
\mu_{\rm B} &=& \left[1+e^{-\eta\alpha^2(1-\cos\theta)}\right]^{1/2}/\sqrt{2}
,\\
\label{qubusbasis2b}
\mu_{\rm E} &=& \left[1+e^{-(1-\eta)\alpha^2(1-\cos\theta)}\right]^{1/2}/\sqrt{2}\,,
\end{eqnarray}
and $\xi\equiv\alpha^2\sin\theta$.
Here and in the following, we assume $\alpha$ to be real.
Now tracing over the loss mode and using the ``A'' basis,
$\{(|0\rangle_{\rm A}\pm e^{i\xi}|1\rangle_{\rm A})/\sqrt{2}\}$,
as new computational basis,
we can express the ``two-qubit'' density matrix of ``A'' and ``B''
in a very compact way as
\begin{eqnarray}\label{qubitqubus}
\mu_{\rm E}^2|\Phi^+(\mu_{\rm B})\rangle\langle\Phi^+(\mu_{\rm B})|+
(1-\mu^2_{\rm E})|\Psi^+(\mu_{\rm B})\rangle\langle\Psi^+(\mu_{\rm B})|\,.
\end{eqnarray}
The resulting density matrix is a mixture of two nonmaximally entangled states,
\begin{eqnarray}\label{qubitqubus2a}
|\Phi^+(\mu_{\rm B})\rangle&=&\mu_{\rm B}|0\rangle_{\rm A}|u\rangle_{\rm B}+
\sqrt{1-\mu^2_{\rm B}}|1\rangle_{\rm A}|v\rangle_{\rm B},
\\\label{qubitqubus2b}
|\Psi^+(\mu_{\rm B})\rangle&=&\mu_{\rm B}|1\rangle_{\rm A}|u\rangle_{\rm B}+
\sqrt{1-\mu^2_{\rm B}}|0\rangle_{\rm A}|v\rangle_{\rm B}\,.
\end{eqnarray}
The phase-rotated coherent states are now contained in the orthogonal basis
of the corresponding qubus-mode subspace,
\begin{eqnarray}\label{qubitqubus3}
|u\rangle_{\rm B}&=&\frac{1}{2\mu_{\rm B}}
\left(|\sqrt{\eta}\alpha\rangle_{\rm B}+e^{-i\eta\xi}|\sqrt{\eta}\alpha e^{i
\theta}\rangle_{\rm B}\right),\\
|v\rangle_{\rm B}&=&\frac{1}{2\sqrt{1-\mu^2_{\rm B}}}
\left(|\sqrt{\eta}\alpha\rangle_{\rm B}-e^{-i\eta\xi}|\sqrt{\eta}\alpha e^{i
\theta}\rangle_{\rm B}\right)\,.\nonumber
\end{eqnarray}
In the form of Eq.~(\ref{qubitqubus}),
one can easily observe the trade-off of the presence of entanglement
for different photon numbers $\alpha^2$, assuming imperfect transmission,
$\eta < 1$, and reasonable phase shifts, $\theta\sim 10^{-2},10^{-3}$.
Choosing $\alpha$ small means
the density matrix in Eq.~(\ref{qubitqubus}) approaches a pure state,
according to Eq.~(\ref{qubusbasis2b});
however, this pure state is nearly unentangled for too small $\alpha$,
according to Eqs.~(\ref{qubusbasis2a}),(\ref{qubitqubus2a}). Conversely,
for large $\alpha$, we have $\mu_{\rm E}^2 \to 1/2$, leading to an almost
equal mixture of the states of Eqs.~(\ref{qubitqubus2a}),(\ref{qubitqubus2b})
in Eq.~(\ref{qubitqubus}); however, this time, the individual states of
Eqs.~(\ref{qubitqubus2a}),(\ref{qubitqubus2b}) are nearly maximally entangled
Bell states. In other words, the quality of the entanglement is affected either
by the decoherence effect of the channel (for large $\alpha$, when many photons transfer
which-path information into the environment) or by the nonmaximal entanglement
of the pure states in Eqs.~(\ref{qubitqubus2a}),(\ref{qubitqubus2b}) (for small $\alpha$,
when the phase-rotated coherent states are nearly indistinguishable and the initial
atom-light entanglement is weak).

Another interesting feature of the state in Eq.~(\ref{qubitqubus}) is that
we may consider a purification of some copies of it
through local operations on the qubits and the
qubus modes, hence distilling a higher degree of entanglement into a smaller number
of copies. This ``prepurification'' (prior to the
qubus interaction with qubit 2) or ``hybrid entanglement distillation'' will be discussed
in Sec.~\ref{entpurif1}.

\begin{figure}[t]
\includegraphics[width=\columnwidth]{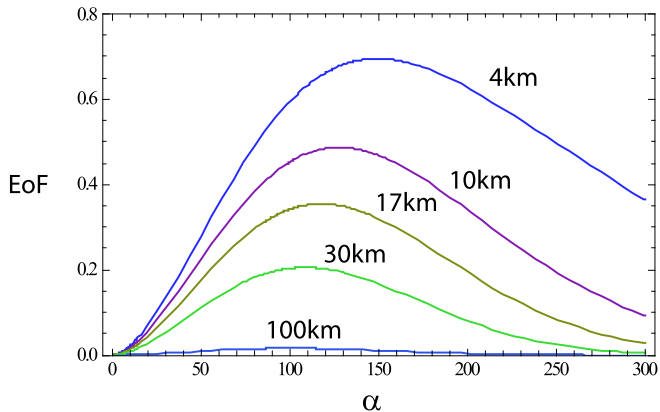}
\caption{\label{fig2} The entanglement of formation of the qubit1-qubus states
as a function of the qubus
amplitude $\alpha$ (square root of qubus photon number) for different channel transmissions, i.e.,
different distances of qubus propagation; photon loss is assumed to be 0.18 dB per km. The phase
$\theta$ is always 0.01.}
\end{figure}

As the density matrix in Eq.~(\ref{qubitqubus}) effectively describes
a two-qubit state, we can evaluate its entanglement
of formation using the concurrence \cite{Wootters98}.
Figure~\ref{fig2} shows the entanglement of formation as a function of the qubus
amplitude $\alpha$ (square root of qubus photon number) for different channel transmissions.
Note that this is the maximum initial entanglement
(prior to any entanglement distillation procedures) available in the repeater protocol.
All the remaining steps of the initial entanglement generation, including the interaction
between qubus and qubit 2 and the measurement of the qubus mode, are local; hence
they can only reduce the amount of entanglement.
The optimal value of the product $\alpha\theta$, maximizing the entanglement, is always of the order
of $\alpha\theta\sim 1$. This is similar to the result obtained for the two-qubit
singlet fidelity {\it after} the interaction with qubit 2 and homodyne measurement of the
qubus mode, reflecting the optimal
trade-off between distinguishability and decoherence \cite{vanloock06}.
Let us now consider the interaction of the qubus mode with qubit 2
and look at the resulting tripartite mixed entangled state.

\subsection{Qubit-qubus-qubit entanglement}\label{entgen2}

In the hybrid quantum repeater protocol,
the entangled state of Eq.~(\ref{qubitqubus}) is subject to a second
interaction, this time between the qubus mode and qubit 2
(which initially is in an equal superposition state).
This interaction results in a controlled rotation of the qubus
by an angle of $-\theta$, as described by Eq.~(\ref{interaction2}) with
$\theta\to-\theta$. The controlled rotation transforms the two orthogonal
qubus basis states of Eq.~(\ref{qubitqubus3}) together with the qubit state as
\begin{eqnarray}\label{2ndinteraction1}
&&|u\rangle_{\rm B}\otimes (|0\rangle_{\rm C}+|1\rangle_{\rm C})/\sqrt{2}
\to\\
&&\quad\quad\quad\quad\quad\quad\quad\quad\quad
\frac{1}{2\sqrt{2}}
\big[|\sqrt{\eta}\alpha\rangle_{\rm B}
(|0\rangle_{\rm C}+e^{-i\eta\xi}|1\rangle_{\rm C})\nonumber\\
&&\quad\quad\quad\quad\quad\quad\quad\quad\quad\quad\quad
+e^{-i\eta\xi}|\sqrt{\eta}\alpha e^{i
\theta}\rangle_{\rm B}|0\rangle_{\rm C}\nonumber\\
&&\quad\quad\quad\quad\quad\quad\quad\quad\quad\quad\quad
+|\sqrt{\eta}\alpha e^{-i
\theta}\rangle_{\rm B}|1\rangle_{\rm C}\big]/\mu_{\rm B},\nonumber
\end{eqnarray}
and, similarly,
\begin{eqnarray}\label{2ndinteraction1}
&&|v\rangle_{\rm B}\otimes (|0\rangle_{\rm C}+|1\rangle_{\rm C})/\sqrt{2}
\to\\
&&\quad\quad\quad\quad\quad\quad\quad\quad\quad
\frac{1}{2\sqrt{2}}
\big[|\sqrt{\eta}\alpha\rangle_{\rm B}
(|0\rangle_{\rm C}-e^{-i\eta\xi}|1\rangle_{\rm C})\nonumber\\
&&\quad\quad\quad\quad\quad\quad\quad\quad\quad\quad\quad
-e^{-i\eta\xi}|\sqrt{\eta}\alpha e^{i
\theta}\rangle_{\rm B}|0\rangle_{\rm C}\nonumber\\
&&\quad\quad\quad\quad\quad\quad\quad\quad\quad\quad\quad
+|\sqrt{\eta}\alpha e^{-i
\theta}\rangle_{\rm B}|1\rangle_{\rm C}\big]/\sqrt{1-\mu^2_{\rm B}}.\nonumber
\end{eqnarray}
Applying these transformations to the density matrix in Eq.~(\ref{qubitqubus}),
a local Hadamard gate to the spin system ``A'',
and a local rotation $e^{i\eta\xi(1-\hat\sigma_z)/2}$ upon system ``C'',
leads to the following tripartite density operator,
\begin{eqnarray}\label{qubitqubusqubit}
\mu_{\rm E}^2|\Phi^+\rangle\langle\Phi^+|+
(1-\mu^2_{\rm E})|\Phi^-\rangle\langle\Phi^-|\,,
\end{eqnarray}
where
\begin{eqnarray}\label{qubitqubusqubit2}
|\Phi^+\rangle&=&\frac{1}{\sqrt{2}}|\sqrt{\eta}\alpha\rangle_{\rm B}
|\phi^+\rangle_{\rm AC}
+\frac{1}{2}e^{-i\eta\xi}|\sqrt{\eta}\alpha e^{i
\theta}\rangle_{\rm B}|10\rangle_{\rm AC}\nonumber\\
&&\quad\quad\quad\quad\quad\quad\quad\quad\,\,
+\frac{1}{2}e^{i\eta\xi}
|\sqrt{\eta}\alpha e^{-i
\theta}\rangle_{\rm B}|01\rangle_{\rm AC},\nonumber\\
|\Phi^-\rangle&=&\frac{1}{\sqrt{2}}|\sqrt{\eta}\alpha\rangle_{\rm B}
|\phi^-\rangle_{\rm AC}
-\frac{1}{2}e^{-i\eta\xi}|\sqrt{\eta}\alpha e^{i
\theta}\rangle_{\rm B}|10\rangle_{\rm AC}\nonumber\\
&&\quad\quad\quad\quad\quad\quad\quad\quad\,\,
+\frac{1}{2}e^{i\eta\xi}
|\sqrt{\eta}\alpha e^{-i
\theta}\rangle_{\rm B}|01\rangle_{\rm AC},\nonumber
\end{eqnarray}
with the maximally entangled Bell states $|\phi^\pm\rangle=
(|00\rangle \pm |11\rangle)/\sqrt{2}$.
The state in Eq.~(\ref{qubitqubusqubit}) with
Eqs.~(\ref{qubitqubusqubit2}) again illustrates the two competing
sources of errors in the entanglement generation step.
Bit-flip errors are caused by the indistinguishability of the phase-rotated
coherent states (if homodyne detection is used for state discrimination);
these can be reduced via sufficiently large photon numbers.
Phase-flip errors occur for any imperfect transmission; this decoherence effect
is suppressed for smaller photon numbers.

In the following, we will consider unambiguous state discrimination (USD)
\cite{IDP1,IDP2,IDP3} of the
corresponding phase-rotated coherent states in Eq.~(\ref{qubitqubusqubit}).
This enables us to completely eliminate bit-flip errors, at the expense of a reduced
efficiency coming from inconclusive measurement results. The relevant, quantum mechanical
USD problem provides ultimate performance bounds. These shall be approached using practical
linear-optics solutions for the required generalized measurements.

\subsection{Conditional state preparation}\label{entgen3}

\begin{figure}[t]
\epsfxsize=\columnwidth \epsfbox[-50 0 160 140]{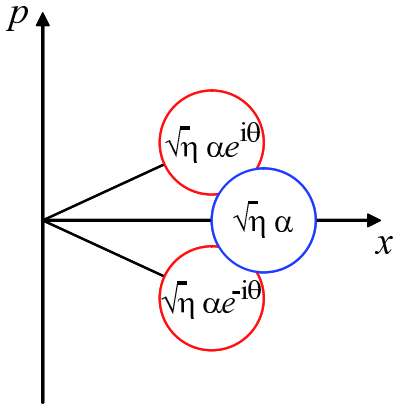}
\caption{\label{fig3} Phase-rotated coherent states to be discriminated
for entangled-state preparation; the unrotated state belongs
to the preferred qubit subspace (even parity); the two rotated states
are correlated with the odd subspace of the two qubits.}
\end{figure}

As illustrated in Fig.~\ref{fig1}a, after the interaction of the qubus with
qubit 2, the final step is to prepare conditionally an entangled two-qubit
state through measurements on the qubus mode including postselection.
The postselection procedure may either filter out approximate, mixed entangled
two-qubit states, still containing some errors from the finite overlap of the phase-rotated
coherent states \cite{vanloock06}; or it may, at a lower succcess rate, perfectly rule out
those contributions belonging to the ``wrong'' coherent states
(see Fig.~\ref{fig3}) and project onto
an entangled two-qubit state whose imperfection originates solely from the losses
in the communication channel. The latter scenario can be achieved via a scheme
based on USD measurements, providing the ultimate, distance-dependent limits on the quality
of the initially generated entangled states. At the same time, it yields
an alternative approach to feasible implementations of the entanglement generation step.
Let us first briefly recall the homodyne-based entanglement generation scheme.

\subsubsection{Homodyne-based state preparation}

A very efficient and practical way to discriminate the phase-rotated
coherent states in Fig.~\ref{fig3} is through homodyne detection \cite{vanloock06}.
While an $x$ measurement could, in principle, project onto both the even and the odd
qubit subspace, a $p$ measurement only leads to an entangled state in the even subspace
and those results consistent with either the $|10\rangle$ or the $|01\rangle$ state
must be discarded. Nonetheless, for the quantum repeater protocol, the $p$ measurement
is preferred to the $x$ measurement, as the errors in the former scale as
$\alpha\theta$ and those of the latter as $\alpha\theta^2$, resulting in fast decoherence
for small $\theta$ and $\alpha$ sufficiently large. Let us now discuss the ultimate
bounds on the performance of the entanglement generation step using an error-free,
USD-based measurement scheme.

\subsubsection{Ultimate bounds}\label{ultimate}

\begin{figure}[b]
\includegraphics[width=\columnwidth]{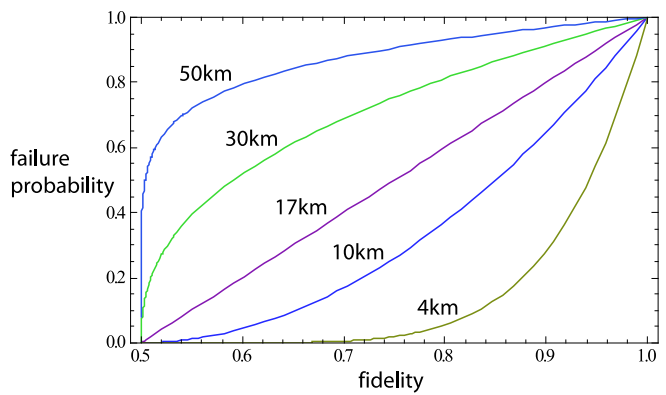}
\caption{\label{fig4} The optimal failure probability for unambiguous
state discrimination of the two density operators in Eq.~(\ref{USD})
as a function of the fidelity of the desired entangled two-qubit state
in Eq.~(\ref{qubitqubusqubit}). The regions below each curve are quantum
mechanically inaccessible. Note that for a transmission $\eta = 1/2$
(17 km), the functional dependence is linear.}
\end{figure}

In order to derive some bounds on the attainable entangled-state
fidelities at a certain rate for a given distance, let us consider the
binary USD problem of discriminating the state $|\sqrt{\eta}\alpha\rangle$
versus the set of states $\{|\sqrt{\eta}\alpha e^{i
\theta}\rangle,|\sqrt{\eta}\alpha e^{-i
\theta}\rangle \}$. We may rephrase this problem as the USD of
the two density operators
\begin{eqnarray}\label{USD}
\hat\rho_1&=&|\sqrt{\eta}\alpha\rangle\langle\sqrt{\eta}\alpha|\,,
\\
\hat\rho_2&=&\frac{1}{2}
\left(|\sqrt{\eta}\alpha e^{i
\theta}\rangle\langle\sqrt{\eta}\alpha e^{i
\theta}| + |\sqrt{\eta}\alpha e^{-i
\theta}\rangle\langle\sqrt{\eta}\alpha e^{-i
\theta}|\right)\,.\nonumber
\end{eqnarray}
Any measurement scheme which is intended to filter out unambiguously
an entangled
Bell state from the two individual states in Eqs.~(\ref{qubitqubusqubit2})
of the mixture in Eq.~(\ref{qubitqubusqubit}) must be a solution
to the above USD problem; thus, the best possible entanglement generation scheme
cannot outperform the optimal USD scheme.
However, note that the converse is not true.
A generalized measurement, even for optimal USD, does not necessarily result
in an entangled state. In particular, unambiguously identifying the state
$\hat\rho_2$ in Eq.~(\ref{USD}) may also mean that the sign of the phase rotation
is determined. In this case, the two qubits end up in a separable state, according
to Eq.~(\ref{qubitqubusqubit}) with Eqs.~(\ref{qubitqubusqubit2}). In fact, in order
to coherently project onto the odd qubit subspace (as it could be done erroneously via
$x$ homodyne detection, see Fig.~\ref{fig3}) through USD, the measurement scheme becomes
less practical involving photon number resolving detectors \cite{Munro05}.

The USD of mixed quantum states is a much more subtle issue than that of pure states
\cite{Rudolph03,Raynal03,Raynal05,Raynal07}. Nonetheless, for equal a priori probabilities
(which is the case we are interested in), the failure probability (the probability for obtaining
an inconclusive measurement outcome) is bounded from below
by the square root of the fidelity of the two
density operators \cite{Raynal05}. For the USD problem of Eq.~(\ref{USD}), this means
\begin{eqnarray}\label{USDbound}
P_? \geq \sqrt{\langle\sqrt{\eta}\alpha|\hat\rho_2|\sqrt{\eta}\alpha\rangle}\,.
\end{eqnarray}
Using Eq.~(\ref{USD}), this leads to the optimal (minimal) failure probability
\begin{eqnarray}\label{USDbound2}
P_?^{\rm opt} = e^{-\eta\alpha^2(1-\cos\theta)}\,.
\end{eqnarray}
This bound can be inserted into the fidelity (this time for the qubit states)
$\mu_{\rm E}^2\equiv F$ of the desired $|\Phi^+\rangle$ state in the mixture of
Eq.~(\ref{qubitqubusqubit}). Using Eq.~(\ref{qubusbasis2b}), we obtain
\begin{eqnarray}\label{USDbound3}
P_?^{\rm opt}(F) = (2 F - 1)^{\eta/(1-\eta)}\,.
\end{eqnarray}
The optimal failure probability as a function of the fidelity
is shown in Fig.~\ref{fig4} for different distances.

The larger the distances,
the larger the failure probabilities become at a given fidelity.
Reasonably high fidelities are only achievable at the expense of small
success probabilities. However, the bound for the USD problem does allow
for fidelities much greater than $1/2$ at distances of 50 km and more.
We will now investigate whether there are practical implementations
of the corresponding USD measurement which approach the quantum mechanical
bounds and hence are no longer limited by distances of 20 km and below.

\subsubsection{Unambiguous state preparation}

Apart from an initial entanglement generation
over potentially larger distances,
there are other advantages of using USD for the conditional
entangled-state preparation. In particular, the resulting
imperfect entangled states will be mixtures of just two
Bell states (rank two mixtures), in the form of Eq.~(\ref{qubitqubusqubit})
after ruling out the odd parity terms in Eqs.~(\ref{qubitqubusqubit2}).
For some copies of this type of mixed-entangled states,
entanglement distillation is more efficient \cite{Zeilinger01}
and the so-called entanglement pumping is no longer bounded by
some fidelity threshold below unity (as for higher rank Bell-diagonal
mixtures) \cite{Duer99}; entanglement pumping means that
spatial resources in the repeater protocol can be turned into temporal
resources by distilling always the same entangled pair with the help
of freshly prepared elementary pairs.

\begin{figure}[b]
\epsfxsize=\columnwidth \epsfbox[0 100 320 210]{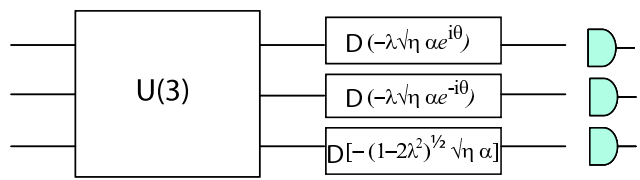}
\caption{\label{fig5} Unambiguous state discrimination of
a zero-phase coherent state from two
phase-rotated coherent states via linear optics, phase-space
displacements, and photon detection.}
\end{figure}

A scheme for unambiguously discriminating the phase-rotated coherent
states in Fig.~\ref{fig3} and hence realizing USD of the density matrices
in Eq.~(\ref{USD}), based upon linear optics and photon detection, is shown
in Fig.~\ref{fig5}. The qubus mode is sent through a linear three-port device,
together with two ancilla vacuum modes, and subsequently, the three output
modes are displaced in phase space before being detected. The three-port
device acts upon a coherent state $|\beta\rangle$ as
\begin{eqnarray}\label{USDlinopt}
|\beta,0,0\rangle \rightarrow
|\lambda\beta,\lambda\beta,\sqrt{1-2\lambda^2}\beta\rangle\,,
\end{eqnarray}
choosing $\lambda$ real. The subsequent phase-space displacements are
\begin{eqnarray}\label{USDlinopt2}
&&\hat D\left(-\lambda\sqrt{\eta}\alpha e^{i
\theta}\right)\otimes
\hat D\left(-\lambda\sqrt{\eta}\alpha e^{-i
\theta}\right)\nonumber\\
&&\quad\quad\quad\otimes
\hat D\left(-\sqrt{1-2\lambda^2}\sqrt{\eta}\alpha\right)
\,.
\end{eqnarray}
Via the three-port device and the displacements,
the three different qubus input states to be discriminated
are transformed as
\begin{eqnarray}\label{USDlinopt3}
&&|\sqrt{\eta}\alpha,0,0\rangle \rightarrow
|\lambda\sqrt{\eta}\alpha (1-e^{i
\theta}),\lambda\sqrt{\eta}\alpha (1-e^{-i
\theta}),0\rangle,
\nonumber\\
&&|\sqrt{\eta}\alpha e^{i
\theta},0,0\rangle \rightarrow\nonumber\\
&&\quad\quad\quad
|0,\lambda\sqrt{\eta}\alpha 2i\sin
\theta),\sqrt{1-2\lambda^2}\sqrt{\eta}\alpha
(e^{i
\theta}-1)\rangle,
\nonumber\\
&&|\sqrt{\eta}\alpha e^{-i
\theta},0,0\rangle \rightarrow\nonumber\\
&&\quad\quad\quad
|-\lambda\sqrt{\eta}\alpha 2i\sin
\theta),0,\sqrt{1-2\lambda^2}\sqrt{\eta}\alpha
(e^{-i
\theta}-1)\rangle.\nonumber\\
\end{eqnarray}

There are now six out of eight possible detection patterns, considering
detectors which do not resolve photon numbers (going
either ``click'' or ``no click''). These patterns unambiguously
identify the corresponding coherent states of the input,
\begin{eqnarray}\label{USDlinopt4}
&&|{\rm click},{\rm click},{\rm no}\, {\rm click}\rangle \rightarrow
|\sqrt{\eta}\alpha\rangle\,,\nonumber\\
&&|{\rm no}\, {\rm click},{\rm no}\, {\rm click},{\rm click}
\rangle \rightarrow
|\sqrt{\eta}\alpha e^{\pm i
\theta}\rangle\,,\nonumber\\
&&|{\rm no}\, {\rm click},{\rm click},{\rm click}
\rangle \rightarrow
|\sqrt{\eta}\alpha e^{i
\theta}\rangle\,,\nonumber\\
&&|{\rm click},{\rm no}\, {\rm click},{\rm click}
\rangle \rightarrow
|\sqrt{\eta}\alpha e^{-i
\theta}\rangle\,,\nonumber\\
&&|{\rm no}\, {\rm click},{\rm click},
{\rm no}\, {\rm click}
\rangle \rightarrow
|\sqrt{\eta}\alpha\rangle
\,{\rm or}\,
|\sqrt{\eta}\alpha e^{i
\theta}\rangle\,,\nonumber\\
&&|{\rm click},{\rm no}\, {\rm click},
{\rm no}\, {\rm click}
\rangle \rightarrow
|\sqrt{\eta}\alpha\rangle
\,{\rm or}\,
|\sqrt{\eta}\alpha e^{-i
\theta}\rangle\,.\nonumber\\
\end{eqnarray}
The pattern $|{\rm no}\, {\rm click},{\rm no}\, {\rm click},
{\rm no}\, {\rm click}\rangle$ is inconclusive (corresponding to the
vacuum contributions from all three modes), whereas the pattern
$|{\rm click},{\rm click},{\rm click}
\rangle$ does not occur at all.

Among the remaining six detection patterns,
there are two patterns which are inconclusive, in addition
to the vacuum-based, inconclusive
pattern $|{\rm no}\, {\rm click},{\rm no}\, {\rm click},
{\rm no}\, {\rm click}\rangle$. The extra inconclusive
patterns are $|{\rm no}\, {\rm click},{\rm click},
{\rm no}\, {\rm click}
\rangle$ and $|{\rm click},{\rm no}\, {\rm click},{\rm no}\, {\rm click}
\rangle$ (the last two patterns of Eq.~(\ref{USDlinopt4})).
As they rule out only $|\sqrt{\eta}\alpha e^{-i
\theta}\rangle$ and $|\sqrt{\eta}\alpha e^{i
\theta}\rangle$, respectively, they are conclusive results
neither for the USD problem nor for the entanglement generation
in the repeater protocol. However, the conditional states
that emerge after the detection of these patterns (i.e., after obtaining
just one click in either the first or the second mode) are, in principle,
still usable for both USD and entanglement generation. Here we will not
consider such conditional dynamics.

Focusing on the remaining four patterns, we observe the following.
Any one of the first four patterns of Eq.~(\ref{USDlinopt4}) conclusively
identifies the quantum state in the USD problem of Eq.~(\ref{USD}).
Among these four, only the first pattern,
$|{\rm click},{\rm click},{\rm no}\, {\rm click}\rangle$, identifies
the state $|\sqrt{\eta}\alpha\rangle$. The other three patterns are only
consistent with the state $\hat\rho_2$ in Eq.~(\ref{USD}), ruling out
$\hat\rho_1$.

If we now look at the entanglement generation step of the repeater
protocol, then even the patterns $|{\rm no}\, {\rm click},{\rm click},{\rm click}
\rangle$ and $|{\rm click},{\rm no}\, {\rm click},{\rm click}
\rangle$ must count as failure, because conclusively identifying
either the state $|\sqrt{\eta}\alpha e^{i
\theta}\rangle$ or the state $|\sqrt{\eta}\alpha e^{-i
\theta}\rangle$ means that the two atomic spins will end up in a separable state,
according to Eq.~(\ref{qubitqubusqubit}) with Eqs.~(\ref{qubitqubusqubit2}).

Eventually, only the two patterns $|{\rm click},{\rm click},{\rm no}\, {\rm click}\rangle$
and $|{\rm no}\, {\rm click},{\rm no}\, {\rm click},{\rm click}
\rangle$ are useful for the entanglement generation. The former one,
conclusively identifying the state $|\sqrt{\eta}\alpha\rangle$, projects
the two qubits onto the even parity subspace. The latter one,
ruling out $|\sqrt{\eta}\alpha\rangle$ and being consistent with both
$|\sqrt{\eta}\alpha e^{i
\theta}\rangle$ and $|\sqrt{\eta}\alpha e^{-i
\theta}\rangle$, leads to the odd subspace. However, in this case,
just obtaining a click for mode 3 is not enough to project the
two qubits onto a maximally entangled state, not even in the ideal case
without losses (see Eq.~(\ref{qubitqubusqubit})). Such a measurement would result
in a superposition of the two ``odd'' Bell states $|\psi^\pm\rangle=
(|10\rangle \pm |01\rangle)/\sqrt{2}$ with an even number (without the vacuum)
and an odd number coherent-state
superposition, $\approx |\alpha i \theta \rangle \pm |-\alpha i \theta \rangle$,
respectively. Thus, only through detection of the photon number parity,
a maximally entangled Bell state of the two qubits can emerge.
This could be achieved via photon number resolving detectors \cite{Munro05}.

As a result, we obtain a highly practical solution for entanglement generation,
based upon two detectors firing at the same time,
$|{\rm click},{\rm click},{\rm no}\, {\rm click}\rangle$.
We denote the success probability for this event to occur as
$P^{\rm even}$. Similarly, the probability for projecting onto
the odd subspace shall be $P^{\rm odd,USD}$ and $P^{\rm odd,ent}$,
where $P^{\rm odd,USD}$ includes those patterns which may or may not resolve the two
states $|\sqrt{\eta}\alpha e^{i
\theta}\rangle$ and $|\sqrt{\eta}\alpha e^{-i
\theta}\rangle$ and hence are only partly useful for entanglement generation
(but still entirely for USD). The probability $P^{\rm odd,ent}$ only includes the pattern
$|{\rm no}\, {\rm click},{\rm no}\, {\rm click},{\rm click}
\rangle$, which, using photon number resolving detectors,
leads to an entangled state. For these probabilities, we obtain,
\begin{eqnarray}\label{probs}
P^{\rm even}&=&\frac{1}{2}\left(
1 - e^{-\lambda^2\eta\alpha^2 \,2 (1-\cos\theta)}\right)^2\,,
\\
P^{\rm odd,USD}&=&
\frac{1}{2}\left(
1 - e^{-(1-2\lambda^2)\eta\alpha^2 \,2 (1-\cos\theta)}\right)\,,
\nonumber\\
P^{\rm odd,ent}&=&
P^{\rm odd,USD}\,\,\times\,\,
e^{-\lambda^2\eta\alpha^2 \,4 \sin^2\theta}\,.
\nonumber\\
\end{eqnarray}

Finally, we use
$P^{\rm total,USD}=P^{\rm even}+P^{\rm odd,USD}$ and
$P^{\rm total,ent}=P^{\rm even}+P^{\rm odd,ent}$ to describe the corresponding
total success probabilities.

\begin{figure}[t]
\includegraphics[width=\columnwidth]{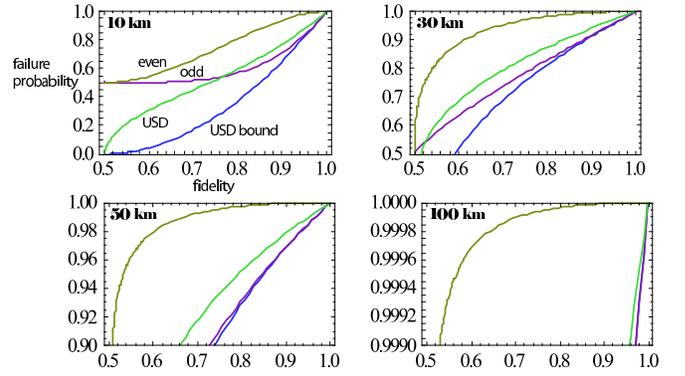}
\caption{\label{figPLOTSARRAY} Failure probabilities as functions of the final two-qubit
maximally entangled-state fidelities for different distances. The regions below ``USD bound''
are quantum mechanically inaccessible. The curves ``USD'' correspond to those
linear-optics implementations in which
all conclusive patterns for both even and odd subspaces are combined (choosing a beam splitter
parameter $\lambda = 0.4$). The plots for ``even'' ($\lambda = 0.7$)
and ``odd'' ($\lambda = 0.01$) describe those measurement
schemes where only a single detection pattern is used in order to project onto
the respective two-qubit subspaces ($|{\rm click},{\rm click},{\rm no}\, {\rm click}\rangle$
for ``even'' and $|{\rm no}\, {\rm click},{\rm no}\, {\rm click},{\rm click}
\rangle$ for ``odd'').}
\end{figure}

Compared to the ultimate bounds derived in
Sec.~\ref{ultimate}, we may now consider three different scenarios.
First, the most practical scheme for entanglement generation, namely, by
unambiguously identifying the state $|\sqrt{\eta}\alpha\rangle$, projecting
the two qubits onto a mixture of even-parity entangled Bell states.
This scheme works with a probability of $P^{\rm even}$ (which is always smaller
than 1/2), and does not require photon number resolving detectors.
Secondly, we consider the odd qubit subspace for entanglement generation,
resulting in a slightly less practical scheme with a need for photon number
resolving detectors; the probability here is $P^{\rm odd,ent}$.
Finally, we add those patterns which resolve the states
$|\sqrt{\eta}\alpha e^{i
\theta}\rangle$ and $|\sqrt{\eta}\alpha e^{-i
\theta}\rangle$ to consider the total probability for a reasonably practical
USD scheme, $P^{\rm total,USD}$. This comparison is shown in Fig.~\ref{figPLOTSARRAY},
where the success probabilities are replaced by failure probabilities,
as functions of the fidelity $\mu_{\rm E}^2\equiv F$
in Eq.~(\ref{qubitqubusqubit}). The linear-optics
parameter $\lambda$ can be used to tune between the even and the odd subspaces.

In Fig.~\ref{figPLOTSARRAY}, again the regions below ``USD bound''
are quantum mechanically inaccessible. The curves ``USD'' correspond to those
linear-optics implementations in which
all conclusive patterns for both even and odd subspaces are combined (choosing a beam splitter
parameter $\lambda = 0.4$); the failure probabilities shown correspond to
$1-P^{\rm total,USD}$.
The plots for ``even'' ($\lambda = 0.7$)
and ``odd'' ($\lambda = 0.01$) describe those measurement
schemes where only a single detection pattern is used in order to project onto
the respective two-qubit subspaces ($|{\rm click},{\rm click},{\rm no}\, {\rm click}\rangle$
for ``even'' and $|{\rm no}\, {\rm click},{\rm no}\, {\rm click},{\rm click}
\rangle$ for ``odd''); the failure probabilities shown correspond to
$1-P^{\rm even}$ and $1-P^{\rm odd,ent}$, respectively.

Here, ``even'' is less efficient, but more practical than ``odd'',
as it does not require photon number resolving detectors. Note that for entanglement generation,
tuning the beam splitter parameter $\lambda$ in order to project onto both even and odd subspaces at the
same time (as for ``USD'' with $\lambda = 0.4$) does not lead to better performances. For larger
distances, also for the case of ``USD'', beam splitter tuning no longer helps;
either projecting onto the even ($\lambda = 0.7$) or the odd ($\lambda = 0.01$)
subspace is optimal in this case as well. Therefore, ``USD'' performs worse than ``odd''
for $30$ km and beyond.

Note that the three patterns of Eq.~(\ref{USDlinopt4}) with clicks
in every mode except one unambiguously identify each individual state
of the set $\{|\sqrt{\eta}\alpha\rangle,|\sqrt{\eta}\alpha e^{i
\theta}\rangle,|\sqrt{\eta}\alpha e^{-i
\theta}\rangle\}$. In other words, for different $\lambda$,
one obtains a family of solutions to the corresponding trinary USD problem.
As the three coherent states here are not symmetrically distributed
($\exp(i\theta \hat a^\dagger \hat a)|\sqrt{\eta}\alpha e^{i
\theta}\rangle\neq |\sqrt{\eta}\alpha e^{-i
\theta}\rangle$ for $\theta\neq 2\pi/3$), the success probability
of the quantum mechanically optimal USD is actually not known.
For realizing optimal USD of $N$ symmetrically distributed coherent states,
protocols have been proposed by van Enk \cite{vanEnk02},
similar to the scheme here,
approximately implementing the optimal $N$-state USD.

\section{Entanglement purification and swapping}\label{entpurif}

In the original hybrid quantum repeater proposal \cite{vanloock06},
the first step is to distribute two-qubit entanglement between
nearest-neighbor stations at a high rate, but with rather modest fidelities.
In order to achieve high-fidelity quantum communication over the
entire repeater channel, the imperfectly entangled qubit states
must be purified; the resulting high-fidelity pairs can be used
to connect the segments of the channel through entanglement swapping.
Further rounds of entanglement purification and swapping will
eventually produce a high-fidelity entangled pair between the remote
ends of the channel \cite{Briegel98,Duer99}.

\begin{figure}[t]
\epsfxsize=\columnwidth \epsfbox[-150 0 400 210]{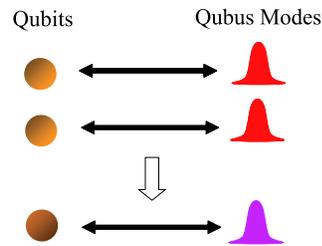}
\caption{\label{fig7} Purifying two copies of a light-matter
hybrid entangled pair into one copy with higher fidelity (purity)
through local operations on the qubits and the light modes.}
\end{figure}

\begin{figure}[t]
\epsfxsize=\columnwidth \epsfbox[-50 0 550 210]{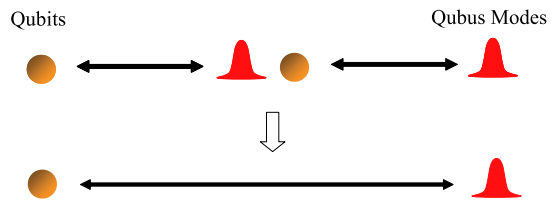}
\caption{\label{fig8} Hybrid entanglement swapping through
unambiguous Bell-state measurements on the joint systems
of qubit and light mode.}
\end{figure}

Instead of using standard entanglement distillation and swapping
procedures on the level of the qubits
\cite{Bennett96,Zukowski,Deutsch,Duer99,Childress1},
we may also consider a purification of the imperfectly entangled,
light-matter hybrid pairs, as described by Eq.~(\ref{qubitqubus}).
In this case, local operations would partly act upon the qubits
and partly on the light field (see Fig.~\ref{fig7}).
This potentially reduces the
number of qubit resources and, moreover, only high-fidelity
entanglement would be transferred from the light modes to the qubits.
Similarly, we could employ a hybrid version of entanglement swapping,
where the Bell-state measurements are performed on the light-matter
hybrid systems (Fig.~\ref{fig8}).

\subsection{Hybrid entanglement distillation}\label{entpurif1}

Entanglement purification of optical, non-hybrid, entangled coherent states
has been considered in Ref.~\cite{jeong02}. Provided the initial copies
of mixed entangled states are of a specific form (rank two mixtures of
a certain pair of entangled coherent, quasi-Bell states), simple linear optics
and photon detectors suffice to enhance the fidelity of the entangled states.
However, for realistic, dissipative environments, the decohered states would not
end up in the desired form; additional local Hadamard-type gates
(transforming Gaussian coherent states into non-Gaussian superpositions of
coherent states) would be needed in order to accomplish the entanglement
purification protocol.

The situation turns out to be similar
for the present hybrid protocol. As can be inferred from Eq.~(\ref{qubitqubus})
with Eqs.~(\ref{qubitqubus2a}),(\ref{qubitqubus2b}), the local Hadamard gates
in an entanglement purification scheme must act upon the qubit states of system
``A'' and the coherent-state superposition basis states of the qubus system ``B'',
Eq.~(\ref{qubitqubus3}). Even though there are recent proposals to achieve
such logical gates for coherent-state superpositions
(``coherent-state quantum computing'' \cite{lund08,ralph03}), using off-line
prepared coherent-state superpositions, the need for extra non-Gaussian, optical
resources (or, possibly, extra CQED-based, optical resources)
may be just as expensive as using additional cavity-based qubit resources.
Therefore we conclude that hybrid entanglement distillation does not
appear to be a more practical alternative to the standard distillation
procedures solely on the level of the two-qubit states.

\subsection{Hybrid entanglement swapping}\label{entpurif2}

In a hybrid version of entanglement swapping,
the Bell-state measurements are performed on the light-matter
hybrid systems (Fig.~\ref{fig8}). For this hybrid Bell measurement,
we can just use the same CQED interactions
as for the initial entanglement distribution, described by Eq.~(\ref{interaction2});
that interaction provides the entangling gate needed for a projection
onto the hybrid ``Bell basis''. A subsequent Hadamard gate can be applied
to the qubit system before measuring both the qubit and the optical qubus
in the ``computational basis''.

More precisely, the following ``Bell states'' are to be discriminated,
\begin{eqnarray}\label{hybridbell}
\left(|0\rangle|\alpha\rangle\pm|1\rangle |\alpha e^{i
\theta}\rangle\right)/\sqrt{2}\,,\nonumber\\
\left(|0\rangle|\alpha e^{i
\theta}\rangle\pm|1\rangle |\alpha\rangle\right)/\sqrt{2}\,.
\end{eqnarray}
In order to distinguish these states, an interaction gate similar to
Eq.~(\ref{interaction2}) is applied, where $|0\rangle|\alpha\rangle\to
|0\rangle|\alpha\rangle$
and $|1\rangle|\alpha\rangle\to|1\rangle|\alpha e^{-i
\theta}\rangle$. The first pair of Bell states in Eq.~(\ref{hybridbell})
is transformed into $\left(|0\rangle\pm|1\rangle\right) |\alpha
\rangle/\sqrt{2}$; in this case, measuring in the Hadamard-rotated
qubit basis reveals the phase of the initial Bell state.
In order to additionally identify the second pair in Eq.~(\ref{hybridbell}),
which will be transformed into $\left(|0\rangle|\alpha e^{i
\theta}\rangle\pm|1\rangle |\alpha e^{-i
\theta}\rangle\right)/\sqrt{2}$, apart from the qubit Hadamard gate
and qubit detection, a measurement on the optical qubus mode must discriminate
the unrotated coherent state from the two rotated ones in phase space
(see Fig.~\ref{fig3}).

For a nearly complete Bell measurement (approximately identifying
any one of the four hybrid Bell states), the two rotated coherent states
must not be distinguished by the measurement. This could be achieved via
$x$ homodyne detection. However, as discussed previously, the distinguishability
in phase space scales badly with distance along the $x$ axis. Therefore,
for a partial Bell measurement identifying only half of the Bell states,
either $p$ homodyne detection can be used, or, alternatively, the USD-based
scheme for unambiguously detecting the unrotated coherent state,
as introduced in the preceding sections. In either case, $p$ homodyne or USD
measurement, the efficiency
of the Bell measurement would be limited by $1/2$.

Now using the hybrid Bell measurement for entanglement swapping (Fig.~\ref{fig8})
means projecting subsystems 2 (the first qubus mode) and 3 (the second qubit) of the
initial pair of entangled qubit-qubus states,
\begin{eqnarray}\label{hybridentswap}
\left(|0\rangle|\alpha\rangle_{12}+|1\rangle |\alpha e^{i
\theta}\rangle_{12}\right)
\otimes
\left(|0\rangle|\alpha\rangle_{34}+|1\rangle |\alpha e^{i
\theta}\rangle_{34}\right)\big/2\,,\nonumber
\end{eqnarray}
onto the Bell basis in Eq.~(\ref{hybridbell}).
According to the method described in the preceding paragraph,
the interaction between the qubit (system 3) and the qubus (system 2)
leads to
\begin{eqnarray}\label{hybridentswap2}
&&\big(|0,\alpha,0,\alpha\rangle+|1,\alpha,1,\alpha e^{i
\theta}\rangle\\
&&\quad\quad\quad
+|0,\alpha e^{-i
\theta},1,\alpha e^{i
\theta}\rangle+|1,\alpha e^{i
\theta},0,\alpha\rangle\big)\big/2
\,.\nonumber
\end{eqnarray}
When the first qubus mode (system 2) is unambiguously determined
to be in the state $|\alpha\rangle$, a Hadamard gate on the second qubit
(system 3) plus measurement in the computational basis yields
one of two possible hybrid Bell states for the first qubit
(system 1) and the second qubus mode (system 4), with the phase depending
on the measurement result. In order to obtain any one of the four hybrid Bell
states of Eq.~(\ref{hybridbell}), in addition, the first qubus mode (system 2)
must be coherently projected onto the subspace corresponding to
$\{|\alpha e^{i\theta}\rangle,|\alpha e^{-i\theta}\rangle\}$ (for example,
via USD and photon-number resolving detectors).

In the above entanglement swapping scheme, clearly the most practical choice
is either $p$ homodyne measurement or USD measurement of the unrotated
coherent state. A success probability below $1/2$ does not automatically
render this scheme inferior to the conventional, deterministic entanglement
swapping with the two-qubit entangled states, because in the latter case,
first the entanglement needs to be distributed probabilistically with
a success probability of at most $1/2$ using either $p$ homodyne or USD
measurements. This probabilistic element is now simply incorporated into
the hybrid entanglement swapping protocol. In other words, using the hybrid
Bell-state analysis, two initial entanglement distributions and subsequent
qubit entanglement swapping can be done almost in one go (when the final
hybrid Bell state is again converted into a two-qubit entangled state through
another CQED interaction and selective measurements with probability $1/2$).
However, the Bell-state analysis for the two-qubit entangled states
in the conventional protocol relies upon complicated two-qubit quantum logic gates;
realizable, for instance, using another four CQED-based dispersive interactions
\cite{vanloock06}. The hybrid Bell-state analysis here would not
require any extra dispersive interactions in addition to those for the
initial entanglement distributions. However, a drawback is that we cannot
efficiently purify the hybrid entangled states (as discussed in the
preceding section) in order to combine sequences of hybrid entanglement
swapping steps with hybrid entanglement purification steps.

\section{Conclusion}

In summary, we investigated the protocol for a hybrid
quantum repeater, based upon dispersive light-matter interactions
between electronic spins and bright coherent light,
with respect to the different kinds of entangled states at the
intermediate steps of the protocol and with regard to the
final optical measurements for conditionally preparing
two-qubit entangled states. As an alternative detection scheme,
we propose to apply USD-based measurements on the optical qubus modes.

Compared to the homodyne-based scheme, there are various advantages
of the USD-based protocol. First of all, one source of errors can be
completely eliminated from the protocol, namely those errors
arising from the inability of perfectly discriminating
phase-rotated coherent states in phase space. In the USD scheme,
this imperfection only leads to smaller efficiencies for the entanglement
generation, but the fidelities are unaffected. As a result,
the fidelities are solely degraded through the decoherence effect
caused by photon losses in the communication channel.
By choosing weaker coherent states, the decoherence effect can be
suppressed and, in principle, repeater spacings of far beyond 10 km
are possible at the expense of smaller entanglement distribution
rates. For example, initial fidelities of about 0.7 are achievable
over 50 km and 100 km with success probabilities of about 1\% and 0.01\%,
respectively, using simple on-off detectors (discriminating between
vacuum and non-vacuum states).
The final two-qubit entangled states here, being, in principle,
ideal rank two mixtures, can be purified very efficiently.
For the USD-based protocol, we also derived ultimate, distance-dependent bounds on
the performance of the entanglement generation step in terms
of success probabilities and fidelities.

Finally, we examined the entanglement purification and swapping steps
for the hybrid repeater protocol from a different perspective.
Instead of performing these steps solely on the
level of the two-qubit entangled states, we considered purification and swapping
with the hybrid entangled states of the atomic qubit and the optical
qubus mode. It turns out that entanglement purification is difficult to achieve,
unless optical, non-Gaussian gates (such as Hadamard gates acting upon
coherent-state superposition states) are available. Hybrid entanglement swapping,
however, can be accomplished easily with exactly the same resources as used
for the initial entanglement distribution. In fact, the probabilistic
entanglement distribution steps can be incorporated into the hybrid entanglement
swapping step, leading to the same overall efficiencies as for the deterministic,
qubit entanglement swapping requiring complicated, less feasible quantum logic gates.
However, a combination of hybrid entanglement swapping with hybrid entanglement
purification in a nested repeater protocol
would again require optical, non-Gaussian gates.

\acknowledgments

PvL acknowledges support from NICT in Japan and the
Emmy Noether programme of the DFG in Germany.
WJM acknowledges the EU programme QAP.
KN acknowledges support in part from NICT and MEXT
through a Grant-in-Aid for Scientific Research on
Priority Area ``Deepening and Expansion of Statistical
Mechanical Informatics (DEX-SMI),'' No. 18079014.
NL acknowledges QAP, QuantumWorks, and the Ontario Centres of Excellence.

\end{document}